\documentclass[
  a4paper,
  english,
  reprint,
  superscriptaddress
]{revtex4-1}

\usepackage[T1]{fontenc}
\usepackage[utf8]{inputenc}
\usepackage{amsmath}
\usepackage{amssymb}
\usepackage{graphicx}
\usepackage{refstyle}
\usepackage[
  citecolor=blue,
  colorlinks,
  linkcolor=blue,
  urlcolor=blue,
]{hyperref}
\usepackage[
  caption=false
]{subfig}

\begin{document}

\title{Transient Uncoupling Induces Synchronization}

\author{Malte Schr\"oder}
\email{malte@nld.ds.mpg.de}
\affiliation{
  Network Dynamics,
  Max Planck Institute for Dynamics and Self-Organization (MPIDS),
  37077 G\"ottingen, Germany
}

\author{Manu Mannattil}
\email{mmanu@iitk.ac.in}
\affiliation{
  Department of Physics,
  Indian Institute of Technology Kanpur,
  U.P. 208016, India
}

\author{Debabrata Dutta}
\thanks{Presently at CGG Services UK}
\email{debabrata.dutta@cgg.com}
\affiliation{
	S.N. Bose National Centre for Basic Sciences,
	Saltlake, Kolkata 700098, India
}

\author{Sagar Chakraborty}
\email{sagarc@iitk.ac.in}
\affiliation{
  Department of Physics,
  Indian Institute of Technology Kanpur,
  U.P. 208016, India
}
\affiliation{
  Mechanics \& Applied Mathematics Group,
  Indian Institute of Technology Kanpur,
  U.P. 208016, India
}

\author{Marc Timme}
\email{timme@nld.ds.mpg.de}
\affiliation{
  Network Dynamics,
  Max Planck Institute for Dynamics and Self-Organization (MPIDS),
  37077 G\"ottingen, Germany
}


\begin{abstract}
  Finding conditions that support synchronization is a fertile and active area of research with
  applications across multiple disciplines.  Here we present and analyze a scheme for synchronizing
 chaotic dynamical systems by transiently \emph{un}coupling them. Specifically, systems coupled only in a fraction of their state space may synchronize even if fully coupled they do not. Although, for many standard systems, coupling strengths need to be bounded to ensure synchrony, transient uncoupling removes this bound and thus enables synchronization in an infinite range of effective coupling strengths. The presented coupling scheme thus opens up the possibility to induce synchrony in (biological or technical) systems whose parameters are fixed and cannot be modified continuously.
\end{abstract}

\maketitle

Synchronization is one of the most prevalent collective phenomena in coupled dynamical systems~\cite{pikovsky03}. Synchronization and related consensus phenomena have been frequently found in biological, ecological, physical, engineering and social systems such as in predator--prey dynamics, the spread of epidemics, the migration of large populations, systems of self-driven
particles and systems of social or technical dynamics~\cite{vicsek95, blasius99, kuperman01, zanette02, wieland10_thesis, ashwin11, strogatz05, matheny14, tanaka14, flunkert10, klinglmayr12}.
For chaotic systems, synchronization typically emerges only within a specific range of coupling strengths and is impossible otherwise~\cite{pikovsky03, pecora97, rosenblum96, timme06}.

In this letter, we propose and analyze a way of inducing synchronization between coupled chaotic oscillators by transient uncoupling: If the system is in a certain predefined
subset of its state space, coupling is active; otherwise it is inactive. 
We systematically study the dependence of successful synchronization on
the fraction of state space where coupling is active. Synchronization may emerge even for systems that coupled continuously in time (i.e., standard coupling) do not synchronize.  Furthermore,  the system may synchronize for an infinite range of coupling strengths, although this is often not possible for ordinarily coupled chaotic systems. 
A systematic numerical analysis reveals how transverse stability properties vary across the attractor with the location of active coupling, not only between more or less stable synchrony but all the way from stability to instability for the same system. This demonstrates that transient uncoupling modifies the collective dynamics in a non-trivial way. These results may find applications in inducing synchrony in systems whose local coupling parameters cannot be continuously varied with ease but only switched on or off.\\

\textbf{Standard coupling.} To start, consider a system of two unidirectionally coupled chaotic oscillators
\begin{eqnarray}
  \label{eq:master}
  \frac{d\mathbf{x}_1}{dt} &=& \mathbf{F}(\mathbf{x}_1), \\
  \label{eq:slave}
  \frac{d\mathbf{x}_2}{dt} &=& \mathbf{F}(\mathbf{x}_2) + \alpha \mathbf{C} \times (\mathbf{x}_1 - \mathbf{x}_2)\, ,
\end{eqnarray}
where $\mathbf{x}_1(t), \mathbf{x}_2(t) \in \mathbb{R}^d$ denote the states of the driving and driven unit, respectively, $\mathbf{C}$ is a square coupling matrix, and $\alpha$ the coupling constant which determines the overall strength of coupling~\cite{pecora97}. As an explicit example throughout this Letter we consider identical $x$-coupled R\"ossler oscillators defined by
$\mathbf{F}(\mathbf{x}) = \left(-(y + z), x + ay, b + z(x - c)\right)^\mathsf{T}$ \cite{roessler76}
and $\mathbf{C} \in \mathbb{R}^{3 \times 3}$, where $\mathbf{C}_{ij} = 1$ for $i=j=1$ and $\mathbf{C}_{ij} = 0$ otherwise.
Further, $a = b = 0.2$, $c = 5.7$ and we take $\mathbf{x}_i=:(x_i,y_i,z_i)^\mathsf{T}$ as a convenient notation.
Other chaotic systems exhibit qualitatively the same phenomena as those presented below \cite{supplement}\nocite{pecora98,lorenz63,chen99}.

\begin{figure}
    \includegraphics[width=0.5\textwidth]{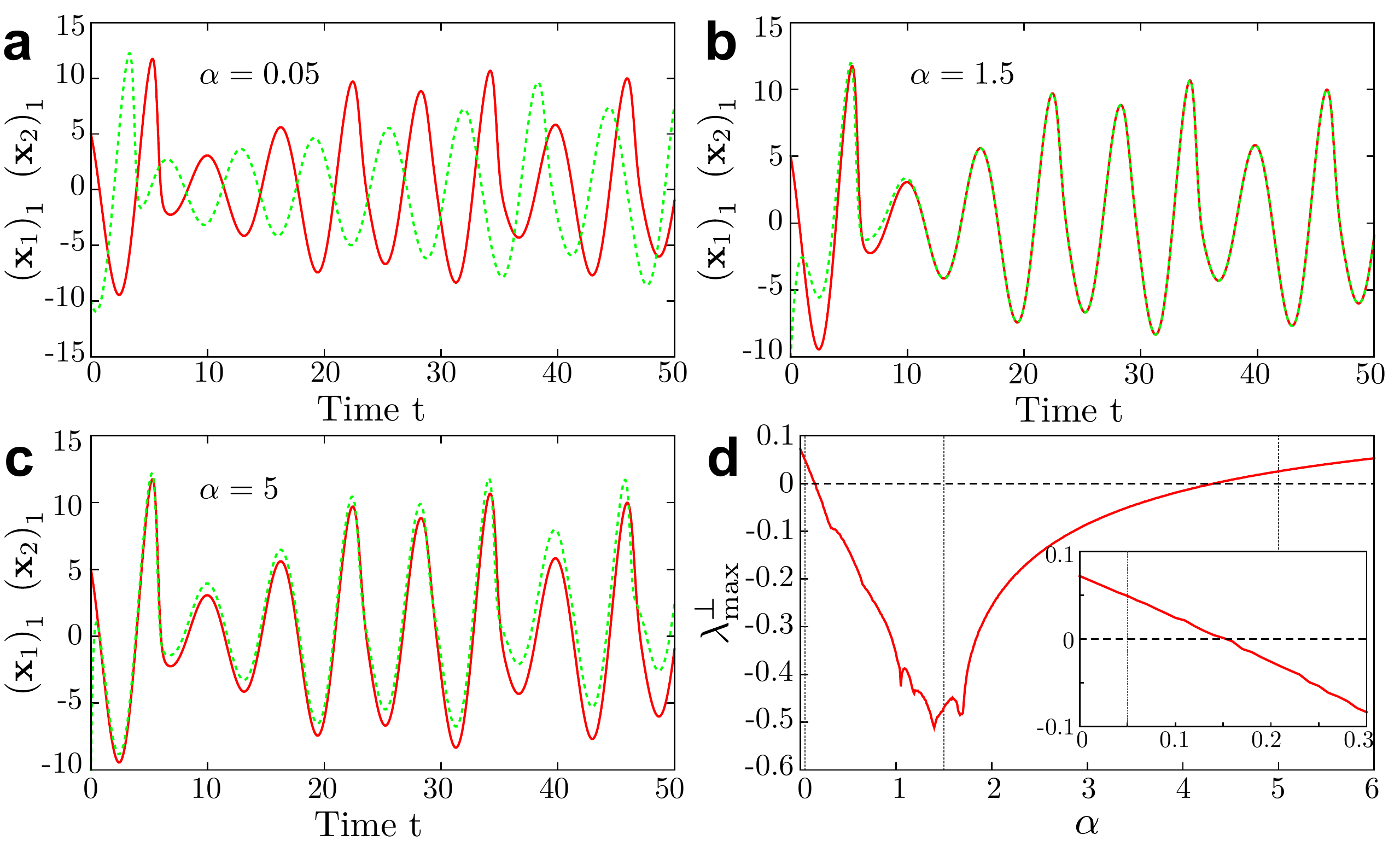}
  \caption{
    \label{fig:lyap}
    Synchronization depends on coupling strength. Trajectories of the driving (solid line) and driven (dashed line) unit of two coupled chaotic oscillators for (a) $\alpha=0.05$, (b) $\alpha=1.5$, and (c) $\alpha=5$ as indicated in panel (d). (d) The maximum transverse Lyapunov exponent $\lambda_\text{max}^\perp$ indicates synchronization for intermediate coupling only.
  }
\end{figure}

Depending on the coupling strength $\alpha$, such systems do or do not synchronize towards  $\mathbf{x}_1(t) = \mathbf{x}_2(t) =: \mathbf{x}_S(t)$. In particular, like many other coupled chaotic systems, R\"ossler oscillators are known to typically synchronize for intermediate coupling strengths $\alpha$, but not if coupled too strongly or too weakly  (Fig.~\ref{fig:lyap}a--c). 

These qualitative synchronization properties depend on the (`transverse') dynamics of the  difference $\mathbf{x_{\bot}} = \mathbf{x}_1 - \mathbf{x}_2$.
A Taylor expansion to first order in the $\left(\mathbf{x_{\bot}}\right)_i$
yields
\begin{eqnarray}
    \mathbf{\dot{x}_{\bot}} &=& \mathbf{F}(\mathbf{x}_1) - \mathbf{F} (\mathbf{x}_2) - \alpha\mathbf{C} \times (\mathbf{x}_1 - \mathbf{x}_2) \nonumber\\
                            &\approx & [\mathbf{J}(\mathbf{x}_S(t)) - \alpha\mathbf{C}]\mathbf{x}_{\bot} \label{eq:varcoup}
\end{eqnarray}
where $\mathbf{J(x)} = \partial_\mathbf{x}\mathbf{F}(\mathbf{x})$ is the local Jacobian of $\mathbf{F}$. For the system~(\ref{eq:varcoup}) to relax to  $\mathbf{x}_{\bot}(t)\rightarrow 0$, its maximum transverse Lyapunov exponent 
\begin{equation}
\lambda_\text{max}^\perp = \lim_{t\rightarrow\infty} \frac{1}{t}\ln \frac{\left|\mathbf{x}_{\bot}(t)\right|}{\left|\mathbf{x}_{\bot}(0)\right|}
\end{equation}
 needs to be negative \cite{pikovsky03}. Fig.~\ref{fig:lyap}d illustrates $\lambda_\text{max}^\perp$ as a function of the coupling constant $\alpha$.  This clearly links in a quantitative way the coupling strength and the qualitative changes in collective dynamics observed before (Fig.~\ref{fig:lyap}a--c).\\

\textbf{Transient uncoupling.} We now introduce transient uncoupling via a factor
\begin{equation} \label{eq:indicator}
  \chi_{A}(\mathbf{x}_2) =
    \begin{cases}
      1  &  \textrm{ for } \mathbf{x}_2 \in A \,;\\
      0  &  \textrm{ for } \mathbf{x}_2 \notin A \, ,
    \end{cases}
\end{equation}
in the coupling term,
\begin{equation}
  \frac{d\mathbf{x}_2}{dt} = \mathbf{F}(\mathbf{x}_2) + 
\alpha \chi_{A}(\mathbf{x}_2) \mathbf{C} \times (\mathbf{x}_1 - \mathbf{x}_2)\, .
\label{eq:clip_slave}
\end{equation}
Here $A \subseteq \mathbb{R}^d$ is a subset of the driven unit's state space where coupling is active.
The two units are thus effectively coupled only within a subset $\mathbb{R}^d\times A$ of their common state space. 
For $A=\mathbb{R}^d$, the units are ordinarily coupled continuously in time. 

\begin{figure}
  \includegraphics[width=0.45\textwidth]{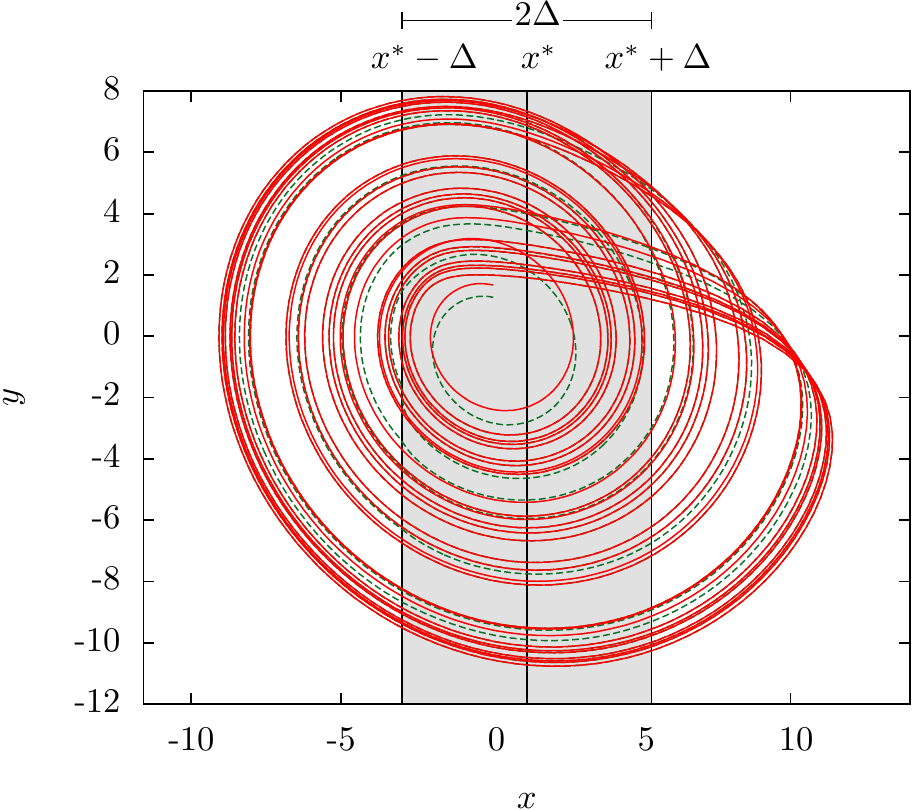}
  \caption{
    \label{fig:clipping}
    Transient uncoupling through state space clipping. The dynamics of two synchronized chaotic oscillators in the $x$-$y$ plane with
    $x^* = (\mathbf{x}^*_2)_1= 1.20$, $\Delta = 4.16$, and $\alpha = 7.0$ (driving: solid curve; driven: dashed curve).  Coupling is only active in the interval  $x_2 \in [ x^* - \Delta  , x^* + \Delta ] $ (shaded in gray).
  }
\end{figure}

Practically relevant subsets $A$ are defined by clipping a region of state space along the direction of a particular coordinate axis,
\begin{equation}
  \label{eq:symmetric_A}
  A_\Delta = \{\mathbf{x}_2 \in \mathbb{R}^{d} : | (\mathbf{x}_2)_1 - (\mathbf{x}^*_2)_1 | \leq \Delta \}\,,
\end{equation}
where $\mathbf{x}^*_2$ is a suitable point and the subscript $'1'$ refers to the first coordinate
of $\mathbf{x}_2$ and $\mathbf{x}^*_2$. Thus, coupling is only active within a column of width $2\Delta$ centered around $(\mathbf{x}^*_2)_1$.
Here, $(\mathbf{x}^*_2)_1 = 1.2$ was chosen as the center of the attractor in $x$-direction.
An example realization for R\"ossler oscillators is illustrated in Fig.~\ref{fig:clipping}.

\begin{figure}
    \includegraphics[width=0.45\textwidth]{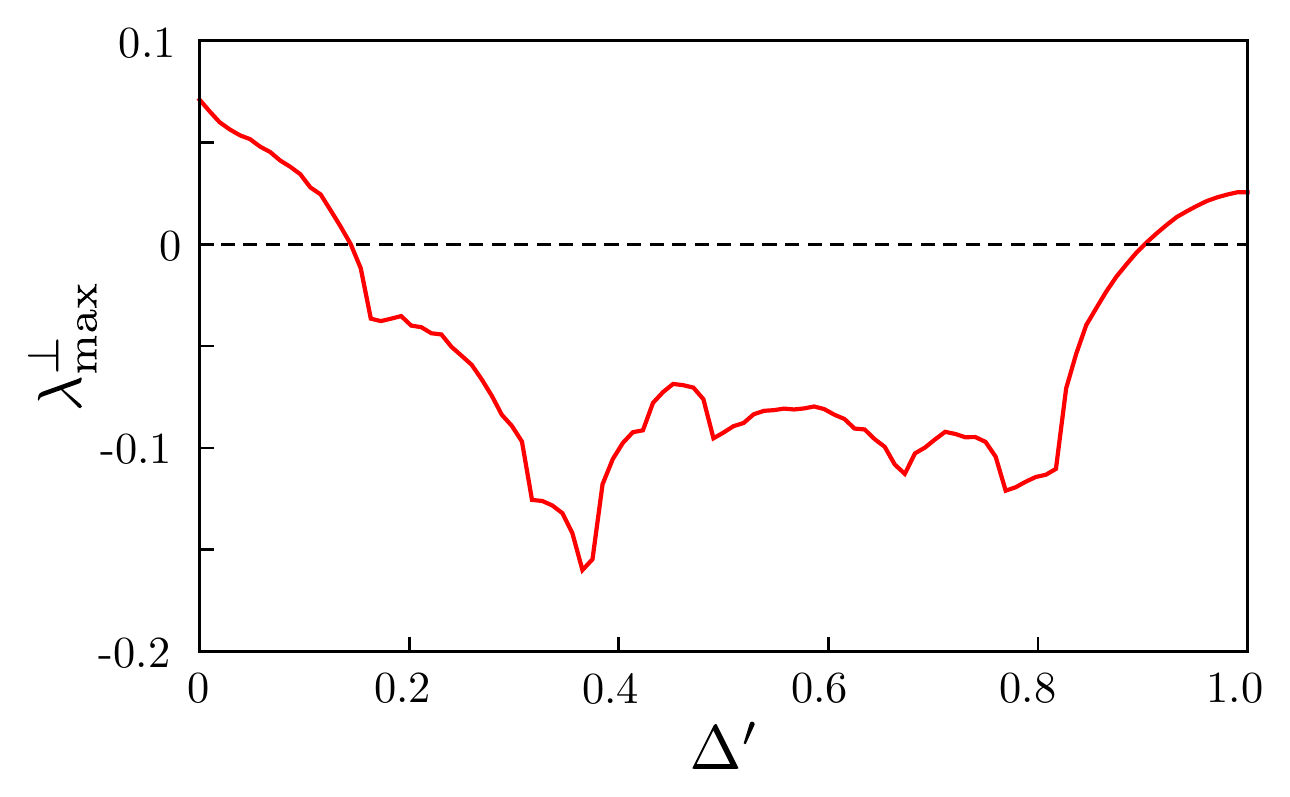}
  \caption{
    \label{fig:clipping_lyap_delta}
    Synchronization induced by transient uncoupling. Maximum transverse Lyapunov exponent for two transiently uncoupled chaotic oscillators (parameters see text) for $\alpha=5$ and clipping with $(\mathbf{x}^*_2)_1 = 1.20$. Synchronization emerges for moderate clipping, i.e., intermediate values of $\Delta '$, although not without clipping ($\Delta' = 1$).
  }
\end{figure}

\begin{figure}
    \includegraphics[width=0.5\textwidth]{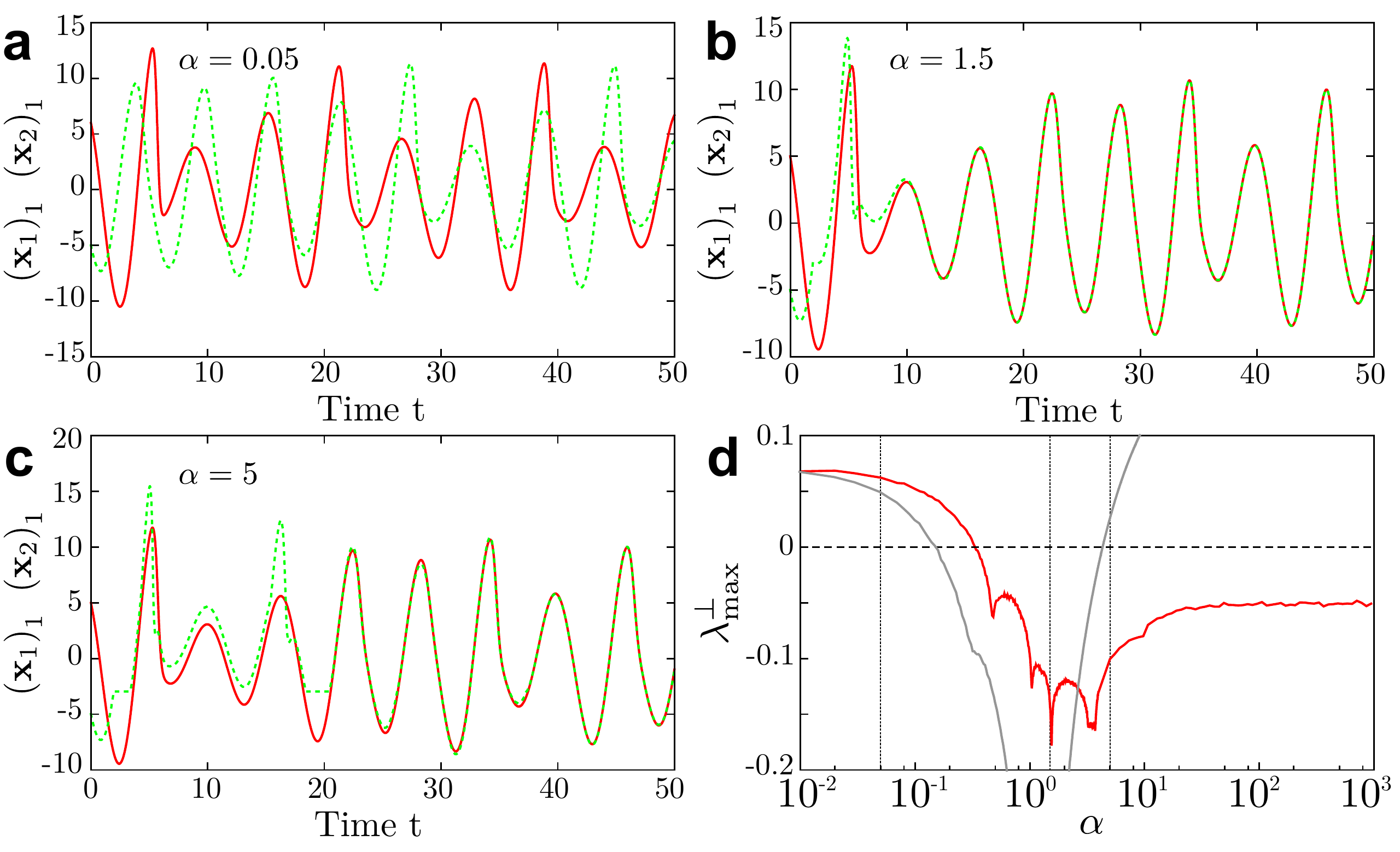}
  \caption{
    \label{fig:clipping_lyap} 
    Transient uncoupling induces synchronization in an infinite range of coupling strengths. Trajectories of the driving (solid line) and driven (dashed line) units for (a) $\alpha=0.05$, (b) $\alpha=1.5$, and (c) $\alpha=5$, the same as in Fig.~\ref{fig:lyap}. The clipping is given by Eq.~\ref{eq:indicator} and \ref{eq:symmetric_A} with $(\mathbf{x}^*_2)_1 = 1.20$ and $\Delta = 4.16$ as in Fig.~\ref{fig:clipping}. (d) Maximum transverse Lyapunov exponent $\lambda_\text{max}^\perp$ as a function of the coupling strength $\alpha$, note the logarithmic scale. The grey line shows $\lambda_\text{max}^\perp$ for normal, unclipped coupling. With transient uncoupling synchronization is stable for arbitrarily large coupling strengths.
  }
\end{figure}

Such transient uncoupling modifies the collective dynamics of the coupled system in a non-trivial way (Fig.~\ref{fig:clipping_lyap_delta} and \ref{fig:clipping_lyap}). Specifically, for a fixed coupling strength $\alpha$, for which standard coupling would not lead to synchronization, clipping in an intermediate interval $\Delta$ induces synchronization. Obviously, for $\Delta \rightarrow 0$ the units become completely uncoupled and cannot synchronize. Similarly, for no clipping $\Delta \rightarrow \Omega/2$ (where $\Omega$ is the width of the attractor along the clipping direction) we reobtain the original system with standard coupling which does not synchronize. For intermediate clipping, however, we find stable synchronization. As the clipping fraction $\Delta$ becomes just one additional parameter of the system we expect the Lyapunov exponent to vary continuously with respect to $\Delta$.
An analysis of the transverse Lyapunov exponent as a function of the clipping fraction $\Delta ' = 2\Delta / \Omega$ confirms this (Fig.~\ref{fig:clipping_lyap_delta}).

Intriguingly, we find that for a fixed clipping interval $\Delta$, the dependence on the coupling strength $\alpha$ is changed not only quantitatively but also qualitatively (compare Fig.~\ref{fig:clipping_lyap} to Fig.~\ref{fig:lyap}). In particular, for intermediate transient uncoupling (intermediate values of $\Delta$) synchrony emerges in an infinite range of coupling strengths $\alpha$, thus in particular for arbitrarily large coupling (Fig.~\ref{fig:clipping_lyap}d). This is in contrast to many chaotic oscillators which, when ordinarily coupled, exhibit an upper bound above which synchronization fails \cite{pecora97}.
In fact we explicitly checked that the same phenomenon also emerges in R\"ossler oscillators with other parameters and in pairs of coupled Lorenz and coupled Chen oscillators as well as for larger networks \cite{supplement}.\\

\begin{figure}
  \includegraphics[width=0.5\textwidth]{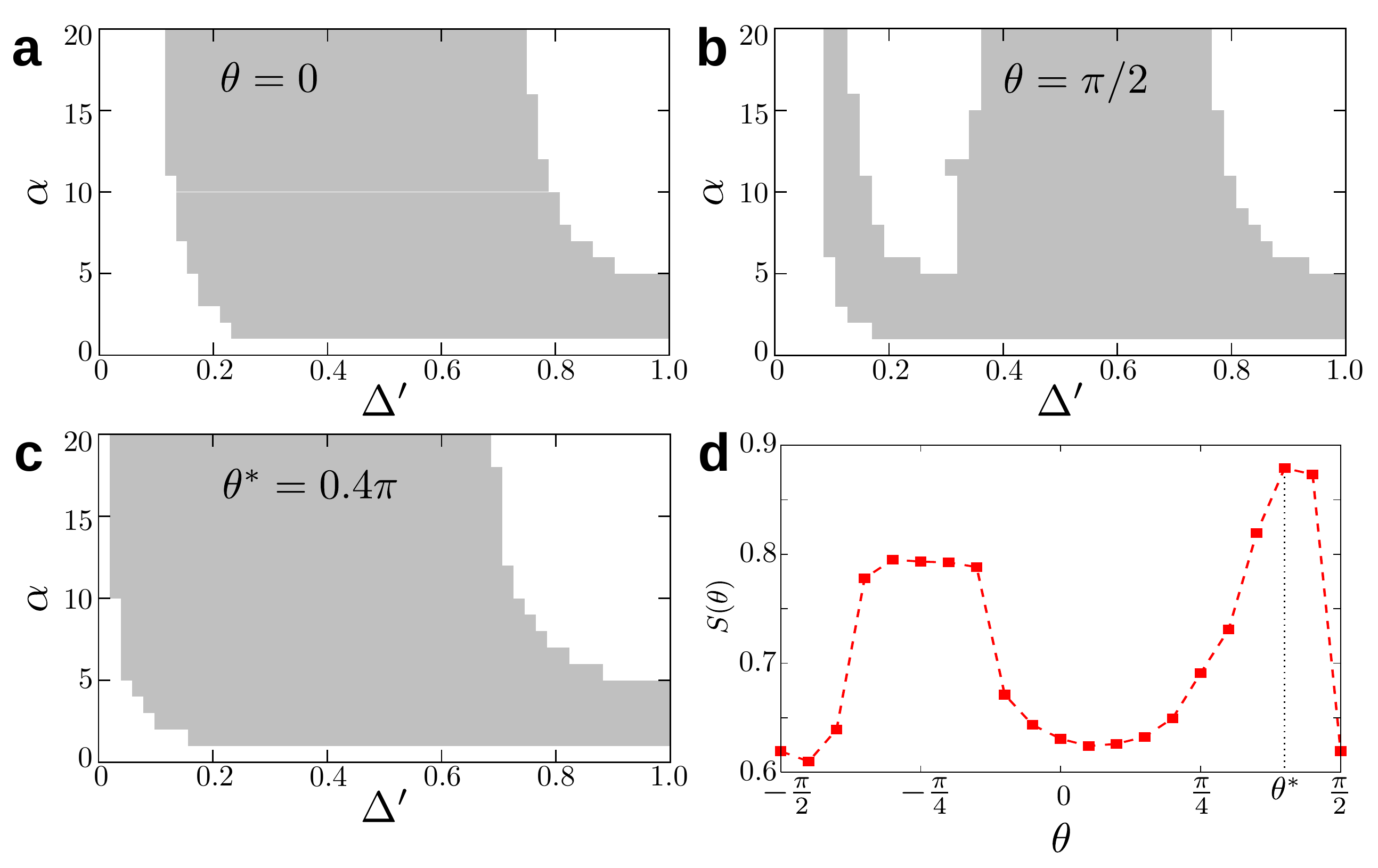}
  \caption{
   \label{fig:rossheatmap}
    Extended synchronization range by transient uncoupling and optimal clipping. (a-c) Depending on the coupling strength $\alpha$ and the percentage ($\Delta ' = 2\Delta/\Omega$) of the state space
    where coupling is active, the system may or may not synchronize. The dark area marks the parameters where the synchronized state is stable, i.e., $\lambda_\text{max}^\perp < 0$. Clipping is (a) in $x$ direction ($x^* = (\mathbf{x}^*_2)_1 = 1.2$), (b)
    in $y$ direction ($y^* = (\mathbf{x}^*_2)_2 = -1.5$), and (c)
    in the direction $y \approx 3.1x$ ($\theta^* = 0.4 \pi$).  
    Accordingly, the direction of clipping can be optimized to achieve the largest possible clipping range. (d) Effectiveness $S(\theta)$ (Eq.~\ref{eq:coup_eff}) of clipping along the direction $\theta$ for fixed $\alpha = 10$.
  }
\end{figure}

\textbf{Optimal uncoupling.} 
We now analyze direction dependencies of transient uncoupling.
Interestingly, the range of coupling strengths $\alpha$ for which the system synchronizes increases when the clipping fraction decreases from $\Delta '=1$, as Fig.~\ref{fig:rossheatmap}a illustrates.  Moreover, the range of clipping fractions for which synchronization emerges depends on the exact direction in state space along which clipping is applied. For instance, clipping along the $x$-axis seems more synchronizing in this sense than clipping along the $y$-axis (compare Fig.~\ref{fig:rossheatmap}a to Fig.~\ref{fig:rossheatmap}b). Oblique directions exhibit even broader ranges of clipping fractions where synchrony emerges (Fig.~\ref{fig:rossheatmap}c).

In fact, certain directions of clipping are optimal.  Due to the shape of the attractor, excursions of trajectories that substantially vary $z$ are rare compared to those that vary the other two coordinates. Thus clipping is desirable in the $x$-$y$ plane. To quantify the effectiveness of clipping depending on its direction in the $x$-$y$ plane, 
we measure the fraction of clipping
\begin{equation}
	\label{eq:coup_eff}
	S\left(\theta\right) = \int_0^1 s\left( f,\theta \right) d f
\end{equation}
for which the system synchronizes when $\alpha$ is fixed. Here, we have measured the angle $\theta$ counterclockwise from the $x$-axis and have defined the synchrony indicator
\begin{equation}
	\label{eq:synch_indicator}
	s\left( f,\theta \right) =   \begin{cases}
      1  &  \textrm{ for } \lambda_\text{max}^\perp < 0 \,;\\
      0  &  \textrm{ for } \lambda_\text{max}^\perp \geq 0 \,,
    \end{cases}\\
\end{equation}
and the temporal clipping fraction
\begin{equation}
   \label{eq:coupling_fraction}
    f = \lim_{T\rightarrow\infty}\frac{1}{T}\int_0^T \chi_{A}(\mathbf{x}_2(t)) \text{d}t \,,
\end{equation}
such that larger values of $S(\theta)$ indicate that synchronization emerges in a larger range of clipping fractions.

The curve $S(\theta)$ has two local maxima (Fig.~\ref{fig:rossheatmap}d), indicating two locally optimal clipping directions, one of which is globally optimal (at $\theta^* \approx 0.4\pi$).
 Why is there such a complicated dependence on direction?\\

\textbf{Transverse stability depends on uncoupling location.} 
We can better understand the synchronization of the system 
by characterizing the stability for a family of clipping functions $\chi(\mathbf{x})=\chi_A(\mathbf{x})$
where
\begin{equation}
	A = A_{\mathbf{x}_2^*, r}	:= \left\{ \mathbf{x}_2 \in \mathbb{R}^d : \left| \mathbf{x}_2 -\mathbf{x}_2^* \right| \le r\left(\mathbf{x}_2^*\right) \right\} \,.
	\label{eq:clipping_sphere}
\end{equation}
Coupling is thus active if and only if $\mathbf{x}_2$ is in a sphere of radius $r\left(\mathbf{x}_2^*\right)$ around $\mathbf{x}_2^*$. We sample the center points randomly from the
attractor (i.e., the invariant measure) of the uncoupled system and choose the size
$r\left(\mathbf{x}_2^*\right)$ such that the coupling is active during a fraction $f$
of the time.

\begin{figure}
   \includegraphics[width=0.45\textwidth]{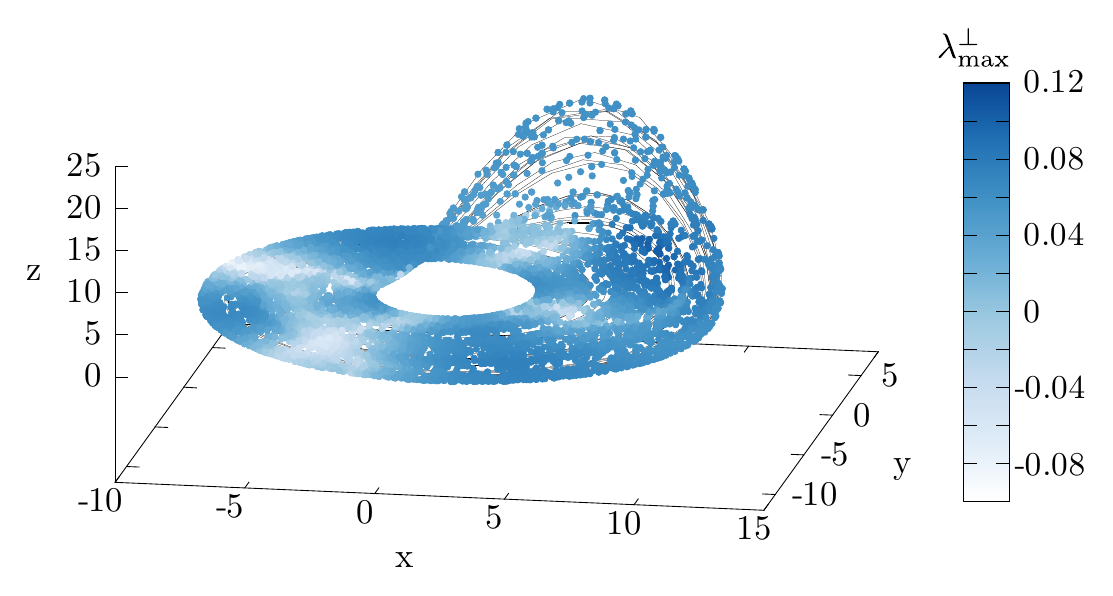}
   \caption{
    \label{fig:lyap_profile_a_5_v_005}
    Multiple switches between stability and instability depending on the coupling location.  The color of the points indicates the maximum transverse Lyapunov exponent of the system 
    with clipping to sets $A=A_{\mathbf{x}_2^*, r}$ (Eq.~\ref{eq:clipping_sphere}), indicating stable synchronization (light) and no synchronization (dark) depending on the coupling location on the attractor. Parameters are $\alpha = 5$ and $f= 0.05$.
   }
\end{figure}

The results show that the impact of the uncoupling strongly depends on the position where clipping is applied. In particular, at identical system parameters, synchrony can be either stable or unstable, depending on where the
coupling is active (Fig.~\ref{fig:lyap_profile_a_5_v_005}). This holds even though the coupling is active for the same fraction $f$ of time.
The attractor regions of positive and negative transverse Lyapunov exponents alternate depending on the direction from the origin.
As these different regions of stability and instability each occur two times on the $2\pi$ phase cycle (circulating the origin) and at roughly equal phase distance,  this explains the two maxima (and the two minima) of the curve $S(\theta)$ found above (Fig.~\ref{fig:rossheatmap}d).
This heterogeneous dependence on the exact location indicates that transient uncoupling, despite being represented by a linear reduction of the coupling term, modifies the collective dynamics of the system in a strongly nonlinear way. As a consequence, the clipping sets $A$ need to be determined individually for each given system to be synchronized.\\

\textbf{Conclusion.} In summary, we have proposed transient uncoupling to modify whether a system of coupled chaotic oscillators synchronizes.  Most generally, these results demonstrate that continuous time coupling is not required for synchronization, even for very simple coupling schemes \cite{junge01}. Interestingly, uncoupling can synchronize systems that would fail to synchronize if ordinarily coupled. Furthermore, it can even remove any upper bound on the coupling strengths that enable synchronization. As a natural extension, it would be challenging to explore how systems capable of weaker forms of collective dynamical coordination, such as phase synchronization, lag synchronization or generalized synchronization would behave if transiently uncoupled \cite{pikovsky03}. Additionally, our scheme may extend synchronization regimes not only in continuous-time systems (described by differential equations, discussed throughout the Letter), but also for chaotic maps and systems temporally switching between different continuous dynamics, cf. e.g.,~\cite{heagy94, amritkar93}.\\

Stability properties of chaotic systems are known to vary locally with the system's state as quantified by the local Lyapunov exponent \cite{abarbanel91, eckhardt93, parlitz14}. For transverse systems, studied above, local stability depends on the direction of the difference vector $\mathbf{x}_{\bot}=\mathbf{x}_1-\mathbf{x}_2$. For small coupling strengths the direction of this vector in the uncoupled transverse system accurately indicates the regions of state space where coupling will be most effective. However, when the coupling is stronger or active in an extended region of state space the trajectories are more strongly modified by the coupling. In particular, whether coupling at one point is effective or not in general depends non-linearly on the coupling in the rest of state space. Optimizing the regions of active coupling in this respect might enhance synchronizability even further.\\

As experimental chaotic systems often exhibit intrinsically fixed or at least restricted internal and coupling settings, the question emerges how to synchronize them. Transient uncoupling by state space clipping may help to induce synchronization for a wider range of coupling strengths, with potential applications to chaotic lasers, electric and electronic circuits, communication systems and chaos based cryptography ~\cite{roy94, sugawara94, sciamanna15, cuomo93, sprott00, bai02, uchida12, annovazzi96, kocarev95, mislovaty03, jiang10}.\\

\begin{acknowledgments}
S.C. gratefully acknowledges financial support from the INSPIRE faculty fellowship awarded by the Department of Science and Technology, Government of India. Partially supported by the Federal Ministry of Education and Research (BMBF) Germany under grant no. 03SF0472E and by a grant from the Max Planck Society, both to MT.
\end{acknowledgments}

\bibliography{manuscript_transientUncoupling_ArXiv}

\end{document}